\documentclass{PoS}

\newcommand{\smgg}{\ensuremath{\mathrm{SU(3)_c} \otimes \mathrm{SU(2)_L} \otimes \mathrm{U(1)}_Y}}
\newcommand{\cgg}{\ensuremath{\mathrm{SU(3)_c}}}

\newcommand{\ewgg}{\ensuremath{\mathrm{SU(2)_L} \otimes \mathrm{U(1)}_Y}}
\newcommand{\wigg}{\ensuremath{\mathrm{SU(2)_L}}}
\newcommand{\ygg}{\ensuremath{\mathrm{U(1)}_Y}}
\newcommand{\emgg}{\ensuremath{\mathrm{U(1)_{EM}}}}

\newcommand{\alphas}{\ensuremath{\alpha_{\mathrm{s}}}}

\newcommand{\chsym}{\ensuremath{\mathrm{SU(2)_L}\otimes \mathrm{SU(2)_R}}}

\newcommand{\m}{\ensuremath{\hbox{ m}}}

\newcommand{\mev}{\ensuremath{\hbox{ MeV}}}

\newcommand{\gev}{\ensuremath{\hbox{ GeV}}}
\newcommand{\tev}{\ensuremath{\hbox{ TeV}}}

\newcommand{\sea}{\mathrm{s}}

\title{DIS and Beyond}

\ShortTitle{DIS and Beyond}

\author{\speaker{Chris Quigg}\\
        Fermi National Accelerator Laboratory\thanks{Operated by Fermi Research Alliance, LLC under Contract No.~DE-AC02-07CH11359 with the United States Department of Energy.}, P.O. Box 500, Batavia, Illinois 60510 USA\\
        E-mail: \email{quigg@fnal.gov}}

\abstract{A digest of my closing overview talk at DIS2013.\hfill {\fbox{\textsf{FERMILAB-CONF-13/225-T}}}}
\FullConference{XXI International Workshop on Deep-Inelastic Scattering and Related Subjects --- DIS2013,\\
		22-26 April 2013\\
		Marseille, France}

\begin{document}

\section{\label{sec:intro}Opening Remarks}
We have enjoyed a rich program of new results on deep-inelastic scattering and related subjects spanning a great range of active topics in particle physics. Viewed from a perspective of a decade or two, the progress is most remarkable. In general terms, in the preceding decades we have formulated, elaborated, and validated two new laws of nature---quantum chromodynamics and the electroweak theory---and established the utility of quarks and leptons as ``elementary'' particles, pointlike at a resolution of $10^{-18}\m$. Many specific results that once seemed the stuff of fanciful dreams are now in hand, and the work in progress---in theory, experiment, and instrumentation---is impressive in its ambition. 

Instead of attempting to distill the excellent Working Group Summaries that we have just heard, I will offer some comments and context on just a few of the central themes of DIS2013. First, I will take a look at the foundational concern of the DIS conference series, nucleon structure, and the window it provides on the strong interaction. Second, I will describe our current understanding of electroweak symmetry breaking after the first wave of experimentation at the Large Hadron Collider. 

\section{\label{sec:proton}What Is a Proton?}
\subsection{\label{subsec:partons}The Parton Model and Beyond}
At high energy, it is fruitful to view a hadron as an unseparated, broadband beam of quarks, antiquarks, and gauge bosons (primarily gluons), and perhaps other constituents, yet unknown. We can look back on forty years of an amazingly robust paradigm, the renormalization-group--improved parton model. Many important processes can be described with high fidelity, starting from three idealizations: the notion that hadron structure and hard-scattering amplitudes factorize, the approximation that parton distributions at a given scale depend only upon the longitudinal momentum fraction, and the simplification that partons are uncorrelated.

Within this framework, confronting measurements with higher-order calculations has led to the development of highly refined parton distributions that describe a plethora of phenomena over a vast kinematic range and begin to exhibit reliable uncertainty estimates. This is a remarkable achievement, the work of many hands, a testament to creativity and perseverance.

I hope that we shall soon be able to better integrate long-distance (low-$Q^2$) and short-distance (large-$Q^2$) views of the nucleon. In particular, studies of elastic and total cross sections at the Large Hadron Collider should lead to improved impact-parameter descriptions of the stuff in a proton. 
A lively conversation with deeply inelastic scattering should ensue!

The one-dimensional idealization is overcome by the development of generalized parton distributions~\cite{gpd},  transverse-momentum distributions~\cite{Collins}, and probes of the three-dimensional structure of the proton~\cite{Boer:2011fh}.
We need also to relax approximations made long ago, for lack of data, such as the universality of sea-quark distributions,
\begin{equation}
u_\sea(x) = \bar{u}_\sea(x) = d_\sea(x) = \bar{d}_\sea(x) = \kappa s(x) = \kappa \bar{s}(x)\;.
\end{equation}
It is true that gluon splitting enforces $q_\sea(x) = \bar{q}_\sea(x)$ perturbatively, but low-scale, nonperturbative processes need not. [We have long known that the ratio $F_{2}^{n}/F_{2}^{p}$ is far from the expectation of $\frac{2}{3}$ that follows from the simple Ansatz $u_{\mathrm{valence}}(x) = 2d_{\mathrm{valence}}(x)$~\cite{Melnitchouk:1995fc}.] 
The chiral quark model, which identifies constituent quarks plus Goldstone bosons as the relevant low-$Q^2$ degrees of freedom, shows how differences between up-quark and down-quark ``sea'' distributions might be established~\cite{Eichten:1992vy}. A constituent quark in the nucleon can fluctuate into a quark plus a pion (or other Goldstone boson). Such reasoning suggests that the observed defect in the Gottfried Sum rule,
\begin{equation}
I_{\mathrm{G}}(Q^2) = \int_0^1 dx \frac{F_2^{p}(x,Q^2) - F_2^{n}(x,Q^2)}{x}\;,
\end{equation}
arises not from antiquarks, but from from the quarks left behind after the emission of a Goldstone boson by a constituent quark. Clinging to ``first approximations'' also misleads when we try to unravel the puzzle of where the nucleon spin resides~\cite{Aidala:2012mv}. 
Chiral-quark-model reasoning suggests that the left-behind quark will have opposite polarization to the constituent quark from which it came, because the Goldstone boson is emitted by a $\gamma_5$-coupling. The quark and antiquark in the Goldstone boson have no net polarization, because the Goldstone boson is spinless. The dissociation entailed in the chiral quark model thus implies that in a proton with positive polarization the quark spin asymetries will be $\Delta d, \Delta s < 0$;  $\Delta\bar{d}, \Delta\bar{s} =0$. An ideal experimental program will measure quark and antiquark polarizations separately.

\subsection{\label{subsec:DM}Nucleon Structure and Dark Matter}
The search for dark matter is a key common interest of particle physics, astronomy, and cosmology~\cite{dmplots}. Searches for weakly-interacting-massive-particle (WIMP) dark matter report limits of observations in terms of the WIMP--nucleon cross section and the WIMP mass. For an important class of supersymmetric dark-matter candidates, the scale of expectations is set by the spin-independent cross section mediated by Higgs-boson exchange~\cite{Baltz:2006fm}. If the Higgs boson interacts with the nucleon principally through its coupling to heavy quark flavors, then pinning down heavy-flavor components of the nucleon is of prime importance~\cite{Thomas:2012tg}. Lattice-QCD calculations are becoming an important complement of the experimental program~\cite{Freeman:2012ry}. The direct searches for WIMPs have attained sensitivities that require calculations of the WIMP--nucleon cross section better than the current uncertainty of a factory of $2\hbox{ - }3$. 

How might we enrich the experimental program already in progress in electron--nucleon scattering? An attractive possibility is a muon-storage-ring Neutrino Factory that could deliver $10^{20}$ $\nu$ per year for on-campus experiments. Such a prodigious flux makes it possible to contemplate neutrino scattering on a polarized target, a hydrogen or deuterium target, or an active target, instead of the massive targets required today. For a sketch of what might be done, see~\cite{Ball:2000qd}. My challenge to the DIS community is to investigate what would constitute the idea suite of neutrino-factory experiments on nucleon structure.

\subsection{\label{subsec:correl}Correlations among the Partons?}
The goodness of the impulse approximation notwithstanding, a parton ``knows'' that it is part of a hadron, particularly at large momentum fractions. In the limit as $x \to 1$, \textit{spin asymmetries} reveal that the SU(6) wave functions---so useful for anticipating static properties of hadrons---do not reflect the spin structure of the nucleon~\cite{Hughes:1983kf}. Evidence for the importance of diquark configurations within hadrons has been in our hands for a long time~\cite{Anselmino:1992vg}. A quark--diquark body plan is apt for doubly-heavy baryons and for high-spin nucleon resonances, to say the least. Bjorken~\cite{Bj2010} has suggested that by studying event structure in high-energy hadron--hadron or lepton--hadron collisions we might be able to see signs of correlations among the partons. In $pp$ collisions at the LHC, diquark--diquark collisions might result in identifiable hot spots.

\subsection{\label{subsec:newphen}New phenomena within QCD?}
The fact that we have not established a solution to the strong CP problem~\cite{sCPp} serves as a reminder that our understanding of QCD is not complete. But unlike the electroweak theory, QCD gives no hint that it could not be self-consistent up to the Planck scale. QCD might nevertheless crack, and we should be attentive to that possibility. At a mild level, we may encounter situations in which factorization breaks down, compromising our ability to make reliable perturbative calculations. At an intermediate level of surprise, we could see evidence for quark compositeness or discover a larger symmetry that contains \cgg. More dramatically, we might observe free quarks---unconfined color.  

I think it is highly likely that we may find new phenomena \textit{within QCD}. Under special circum\-stances---perhaps in high-multiplicity events---there may be other mechanisms for particle production, beyond the traditional pair of diffraction plus short-range order. The very high density of few-GeV partons in collisions at LHC energies might lead to thermalization. Long-range correlations in rapidity, such as the ``ridge'' observed in $pp$ collisions by the CMS Collaboration~\cite{Khachatryan:2010gv}, may signal new production mechanisms that point to a richer picture of hadron structure than we have commonly assumed for ``soft'' collisions. For example, the collision of aligned flux tubes connecting valence quarks might be manifest as a line source of particles~\cite{Bjorken:2013boa}.

With or without specific motivational conjectures, unusual event structures (observed in informative coordinates) can lead us to uncover new mechanisms and enrich our response to the ``What is a proton?'' question~\cite{Quigg:2010nn}

\section{\label{sec:asf}Asymptotic Freedom}
\subsection{\label{subsec:hadmass} The Origin of Hadron Masses}
In leading logarithmic approximation, the strong coupling $\alphas$ famously evolves~\cite{Gross:1973id} as
\begin{equation}
{\frac{1}{\alphas(Q)} = \frac{1}{\alphas(\mu)} + \frac{(33 - 2n_f)}{6\pi}\ln\left(\frac{Q}{\mu}\right)},
\label{eq:evol}
\end{equation}
where $n_f$ is the number of active quark flavors 
This makes it telling to plot $1/\alphas$ as a function of $\ln{Q}$~\cite{Kronfeld:2010bx}. The scale dependence of $\alphas$ has been computed through four loops~\cite{vanRitbergen:1997va}, but the deviation from a straight line on the semi-logarithmic plot is mild, so the leading-order expression will suffice for the arguments to follow.

The smallness of $\alphas$ at large scales (in practice, already at a few GeV) enables perturbative analysis of the kind familiar in the study of deep-inelastic scattering. In the other limit of small scales, the largeness of $\alphas$ foretells the nonperturbative phenomenon of color confinement. ``Dimensional transmutation'' introduces a dimensionful parameter that sets the scale of hadron masses. Lattice techniques master the nonperturbative problem and allow very impressive---and essentially \textit{a priori}---calculations of the hadron spectrum~\cite{Kronfeld:2012ym}. We have learned from these studies that the nucleon mass is an exemplar of Einstein's insight that mass is related to rest energy through $m = E_0/c^2$. The masses of the up and down quarks directly contribute only a few percent of the nucleon mass:. 
\begin{equation}
3\,\frac{m_u+m_d}{2} = 10 \pm 2\mev .
\end{equation}
The full story is a little more subtle. If we let the quark masses go to zero, holding $\alphas$ fixed, then chiral perturbation theory leads to the conclusion that $M_N \rightarrow 870\mev$~\cite{GaLeu}.
It is a supreme accomplishment of lattice QCD is to show that the quark-confinement origin of nucleon mass  explains nearly all visible mass in the Universe. 

\subsection{\label{subsec:qmass}The Influence of Quark Masses}
Even if the up- and down-quark masses are far too small for the nucleon to follow the Newtonian dictum that the mass of an object is the sum of the masses of the constituents, they shape our world in important ways. The amount by which $m_d > m_u$ is responsible for the inequality $M_p < M_n$ that ensures the stability of the proton against $\beta$-decay. If the value of the strong coupling $\alphas(Q)$ is fixed at some high scale, as would be the case in a unified theory of the strong, weak, and electromagnetic interactions, then the value at the low energies appropriate for the computation of hadron masses is determined by the evolution equation (\ref{eq:evol}), and is thus sensitive to the spectrum of colored particles. 

If we define the QCD scale parameter that arises through dimensional transmutation as 
\begin{equation}
1/{\alphas(2m_c)} \equiv ({27}/{6\pi})\ln\left({2m_c}/{\Lambda}\right)
\end{equation}
and consider variations in the top-quark mass, then it is easy to show~\cite{CQPT} that
\begin{equation}
\frac{\Lambda}{1\gev} = \mathrm{const.}\left(\frac{{m_t}}{{1\gev}}\right)^{2/27}.
\end{equation}
Now, the lattice QCD calculations have taught us that $M_p \approx C\Lambda + \ldots$, where $C$ is computable on the lattice and the neglected terms include quark masses, electromagnetic self-energies, etc.  In this picture, we discover that $M_p \propto m_t^{2/27}$: despite the negligible population of virtual top-antitop pairs in the nucleon, the top-quark mass influences the proton mass. As a small homework assignment, I suggest that you consider how $\alphas$, $\Lambda$, and $M_p$ would change if all the quark masses were set to zero (or very tiny values). If spontaneous breaking of the electroweak symmetry is the origin of quark masses, the Higgs field does not endow the nucleon with mass, but it does influence the value of the mass that results from color confinement---at least in this unified-theory framework. 

\subsection{\label{subsec:unify}A Unified Theory?}
We do not have direct experimental evidence that the \smgg\ interactions are unified into SU(5) or some larger gauge symmetry, but the idea of unification is a natural response to the question, ``Why are atoms so remarkably neutral?''  The different running of the \cgg, \wigg, \ygg\ couplings points to the possibility of coupling-constant unification at a high scale. Joining quarks and leptons in extended families can explain the equality of the proton and positron charges, and also leads to the possibility of proton decay.

According to our current low-energy knowledge of the \smgg\ couplings and the spectrum of particles that carry standard-model charges, The three couplings tend to approach each other at high energies, but the three do not coincide at one point. Perturbative calculations show that coupling-constant unification is more promising in supersymmetric $\mathrm{SU(5)}$ than in the simple $\mathrm{SU(5)}$ theory, provided that the change in evolution due to a full spectrum of superpartners occurs near $1\tev$. Figure~1 of Ref.~\cite{Quigg:2012zm} compares the evolution of $1/\alphas$, in leading logarithmic approximation, with and without a superpartner threshold at $Q = 1\tev$. The slope changes significantly, from $7/2\pi$ to $3/2\pi$. (Corresponding changes in the slopes of $1/\alpha_2$ and $1/\alpha_1$ caused by the richer particle spectrum contribute to the improved coupling-constant unification in this theory.) 

This observation suggests another homework problem---not just a thought experiment, and not easily done. Can we, in LHC experiments, measure the change in slope of $1/\alphas$ that accompanies the onset of supersymmetry? Seeing or not seeing such a change would constitute powerful evidence for or against  a new set of colored particles; it is a global search that complements ongoing searches for specific new-particle signatures. ATLAS and CMS~\cite{ATLAS:2013lla} have begun to probe $\alphas$ at scales approaching $1\tev$ by comparing 3-jet and 2-jet rates. I suspect that $Z + \hbox{jets}$ might provide better control in the long term. It will not be easy to determine $\alphas$ up to say, $3\tev$, and understanding what is measured will demand a continuing dialogue between theory and experiment. But the rewards will be great!

\section{\label{sec:higgs}Electroweak Symmetry Breaking and the Higgs Boson}
\subsection{The Avatar of Electroweak Symmetry Breaking}
Much evidence indicates that the standard-model gauge symmetry is broken---either spontaneously or dynamically---to QCD plus electromagnetism: $\smgg \to \cgg\otimes\emgg$. The LEP experiments gave a lovely demonstration of the secret \ewgg\ symmetry by validating the gauge cancellation among neutrino-, electron-, and $Z^0$-exchange contributions to the reaction $e^+ e^- \to W^+W^-$~\cite{gaugecan}. Many possibilities have been investigated for the hitherto unknown agent that hides electroweak symmetry, including a force of a new character, based on interactions of an elementary scalar;  a new gauge force, perhaps acting on undiscovered constituents;  a residual force that emerges from strong dynamics among electroweak gauge bosons;  and an echo of extra spacetime dimensions. 
 
The discovery of the 126-GeV boson by the ATLAS and CMS collaborations~\cite{Hdisc} indicates that the dominant mechanism for electroweak symmetry breaking is a degenerate vacuum set up by the self-interactions of a multiplet of scalar fields~\cite{Hmech}. The new particle gives every indication of being what has come to be called the standard-model Higgs boson, $H$. Production rates and branching fractions determined by ATLAS~\cite{ATLASHiggs} and CMS~\cite{CMSHiggs} are in good general agreement with standard-model expectations~\cite{HWG}. The LHC  gives us multiple looks at the new particle. The mass of $H$ means that several decay modes are detectable: $\gamma\gamma, WW^*, ZZ^*, \tau^+\tau^-, b\bar{b}$, \ldots. Four production mechanisms---gluon-gluon fusion through the top-quark loop, $H(W,Z)$ associated production, vector-boson fusion, $t\bar{t}H$ production---can be observed.  

\subsection{\label{subsec:questions}Questions about the LHC Higgs Boson}
Is the field made manifest by the Higgs boson $H$ discovered at the Large Hadron Collider the sole source of electroweak symmetry breaking? In other words, does $H$ couple to $W^+W^-$ and $ZZ$ with the expected strength? Does it have companions? A very important branch point is whether the new boson couples also to fermions, because it is a logical possibility that the mechanism of electroweak symmetry breaking is not responsible for fermion mass. We have evidence that $H$ couples to $t\bar{t}$, $b\bar{b}$, and $\tau^+\tau^-$, but still need to verify that the coupling strengths fully account for the fermion masses. We haven't yet shown that $H$ couples to fermions beyond the third generation, although the dimuon decay is a target for the High-Luminosity LHC and lepton-collider Higgs factories would be able to detect charm pairs. Can we ever prove the origin of the electron mass?
  Does $H$ decay into new particles, or through the action of new forces? Are all production modes present at the expected rates? Is $H$ a pure $J^P = 0^+$ state? Will we find any sign of new strong dynamics? What are the implications of $M_H \approx 126\gev$?

\subsection{\label{subsec:matters}Why does discovering the agent matter?}
One way to assess the importance of the new discovery is to imagine a world without a symmetry-breaking (Higgs) mechanism at the electroweak scale. Analyzing the consequences of that \textit{Gedankenexperiment} is rather involved~\cite{Quigg:2009xr}, but we can summarize the main results in a few lines, especially if we consider the restricted case of one generation of quarks and leptons. The electron and the up and down quarks would have no mass. QCD would nevertheless confine quarks into protons, etc. As we have seen in \S\ref{subsec:hadmass}, nucleon mass would be little changed. There is a small surprise: on roughly the energy scale at which confinement occurs, the global \chsym\ chiral symmetry of the massless up and down quarks is spontaneously broken, giving rise to a $\bar{q}q = \bar{q}_{\mathrm{L}}q_{\mathrm{R}} + \bar{q}_{\mathrm{R}}q_{\mathrm{L}}$ condensate. By coupling left-handed and right-handed quarks, which transform differently under \ewgg, the chiral-symmetry-breaking condensate hides the electroweak symmetry, giving tiny masses to $W$, $Z$ (about $2500\times$ less than in our world). If stable species---$\alpha$-particles, for example---were formed in the early universe and survived until late times, they might be candidate nuclei of atoms. But if the electron mass is zero, the Bohr radius of the electron is infinite, so the electron cannot be identified with a particular nucleon. Atoms lose their integrity. No atoms means no chemistry, no stable composite structures like liquids, solids, \ldots no template for life. It is a very different world from ours!

\subsection{\label{subsec:higgsf}Do We Need a Higgs Factory?}
There is no question that, if a Higgs factory that complemented the Large Hadron Collider were in operation today, it would be an important addition to our experimental portfolio. That is not the case: a new Higgs factory would require effort and expense comparable to the LHC, and would not be in operation for more than a decade. Pinning down the mechanism of electroweak symmetry breaking, for which a prime ingredient is forensic Higgs science, is of such importance that the investment and the wait might be worthwhile. I cannot examine the issues thoroughly in a few paragraphs, but I can try to indicate what I see as some of the important considerations.

The Higgs-boson studies that seem most telling at any moment depend on what is already known: whether the particle discovered at the LHC turns out to be very standard-model--like (as the data so far suggest) or deviates significantly from standard-model expectations. The discovery of another ``Higgs-like object'' could change the requirements for what a new Higgs factory must do. Direct evidence for or against new degrees of freedom could similarly influence our assessment of the energy and luminosity needed.

There are many postulants, at different levels of maturity. The best developed is the International Linear Collider, perhaps begun in a Higgs-factory phase at $0.7 \times$ the 500-GeV ILC. A number of circular $e^+e^-$ colliders, limited in energy but possibly with superior luminosity, are discussed. A modest-energy muon collider could in principle be a very refined tool for determining Higgs-boson properties through the formation reaction, $\mu^+\mu^- \to H$. A multi-TeV lepton collider would permit the study of vector-boson fusion and might allow meaningful measurements of the $HHH$ coupling. As we have heard extensively at DIS2013, an electron--proton collider such as the LH$e$C has its own set of potential strengths~\cite{BruceM}. We cannot afford all of these machines; even one will require extraordinary effort.

Due diligence demands that we prepare a reliable shopper's guide. This requires clearly stated assumptions, documented uncertainty estimates, a rich list of observables, including $\Gamma(\mu\mu)$, $M_H$, $\Delta M_H$, $\Delta \Gamma_H$, \ldots, and a rich list of possible machines---including the projected runs of the LHC. I also insist that a time dimension be included. It is easy to be swept away by enthusiasm if we imagine that we could have all the answers next week. The test is, how long will it take \textit{after the machine operates} for us to gather the answers we seek. For the $\ell^+\ell^-$ colliders, we need to assess (under different scenarios, and in view of what the LHC will yield) the importance of collateral measurements of  $M_W$ and of $m_t$, $\alphas$, and the top Yukawa coupling~\cite{Asner:2013hla}, and to weigh the importance of giga-$Z$, or even tera-$Z$ studies. Time spent running at $W^+W^-$ threshold or on the $Z^0$ pole is time not devoted to $HZ$ production.

 \section{\label{sec:unreasonable}The Unreasonable Effectiveness of the Standard Model}
 The successes of the \ewgg\ electroweak theory at tree level---for both charged-current and neutral-current interactions---and as a quantum field theory at one-loop level has elevated it to a law of nature.~\cite{ErlLang,Quigg:2009vq} The quest for new physics should be particularly promising where the standard model predicts very small effects. A leading example is the search for new physics in the flavor sector through the observation of flavor-changing neutral currents~\cite{Branco} at levels very different from the Glashow--Iliopoulos--Maiani-suppressed expectations of the electroweak theory~\cite{Glashow:1970gm}. It is a generic consequence of extensions to the standard model that flavor-changing neutral currents should occur at levels in excess of the standard-model rates, at least in some channels. The search for amplified (or suppressed) decay rates complements direct searches for new phenomena that rely on the production of new, perhaps exotic, particles.
 
 No such evidence has turned up. Extremely telling recent observations concern the rare decay $B_s \to \mu^+\mu^-$, which has been observed by the LHC$b$ and CMS experiments at a branching fraction of ${\cal B}(B^0_{s} \to \mu^+\mu^-) = (2.9 \pm 0.7) \times 10^{-9}$~\cite{Aaij:2013aka}, well compatible with the standard-model expectation, ${\cal B}(B^0_{s} \to \mu^+\mu^-) = (3.56 \pm 0.30) \times 10^{-9}$~\cite{Buras:2012ru}. This is a case in which some classes of supersymmetric models, and other new-physics scenarios, anticipated an important enhancement over the standard-model rate~\cite{Buras:2013uqa}, and so it places important new constraints on those examples.
 
I regard the persistent absence of flavor-changing neutral currents as a strong hint that a new symmetry or new dynamical principle may be implicated, or that new physics is more distant than hierarchy-problem considerations had indicated.

\section*{Acknowledgements and Thanks}
 I am grateful to Cristi Diaconu and the CPPM Team for warm hospitality in Marseille. It is a pleasure to thank all the DIS2013 organizers and participants for a stimulating and lively meeting.


\frenchspacing
\begin{thebibliography}{99}
\bibitem{gpd}
M.~Diehl, Phys.\ Rept.\ \textbf{388} (2009) 41; 
X.~Ji, Ann.\ Rev. Nucl. Part. Sci. \textbf{54} (2004) 413;
A. V. Belitsky and A. V. Radyushkin, Phys. Rept. \textbf{418} (2005) 1.

\bibitem{Collins}
For an overview, see J. C. Collins, \textit{Foundations of Perturbative QCD} (Cambridge University Press, Cambridge, 2011).

\bibitem{Boer:2011fh}
  D.~Boer, {\it et al.},
  ``Gluons and the quark sea at high energies: Distributions, polarization,
  tomography,'' A report on the joint BNL/INT/Jlab program on the science case for an Electron-Ion Collider, arXiv:1108.1713.

\bibitem{Melnitchouk:1995fc}
  W.~Melnitchouk and A.~W.~Thomas,
  Phys.\ Lett.\  {\bf B377}, 11 (1996).

\bibitem{Eichten:1992vy}
  E.~J.~Eichten, I.~Hinchliffe, and C.~Quigg,
  Phys.\ Rev.\ D {\bf 45} (1992) 2269;
  Phys.\ Rev.\ D {\bf 47} (1993) 747.
  
\bibitem{Aidala:2012mv}
  C.~A.~Aidala, S.~D.~Bass, D.~Hasch and G.~K.~Mallot,
  Rev.\ Mod.\ Phys.\  {\bf 85} (2013) 655.
  
\bibitem{dmplots}
For an up-to-date catalogue of dark-matter searches, consult D.~Speller, R.~Gaitskell, and J.~Filippini's  \href{http://cedar.berkeley.edu/plotter/}{DMtools website}.  

  \bibitem{Baltz:2006fm}
  E.~A.~Baltz, M.~Battaglia, M.~E.~Peskin and T.~Wizansky,
  Phys.\ Rev.\ D{\bf 74} (2006) 103521;
  J.~R.~Ellis, K.~A.~Olive and C.~Savage,
  Phys.\ Rev.\ D{\bf 77} (2008) 065026.

\bibitem{Thomas:2012tg}
  A.~W.~Thomas, P.~E.~Shanahan and R.~D.~Young,
  Nuovo Cim.\ C {\bf 035N04} (2012) 3.
  

\bibitem{Freeman:2012ry}  For a recent example, see W.~Freeman and D.~Toussaint  [MILC Collaboration],   ``The intrinsic strangeness and charm of the nucleon using improved staggered fermions,''
  arXiv:1204.3866.
  
  
\bibitem{Ball:2000qd}
  R.~D.~Ball, D.~A.~Harris, and K.~S.~McFarland,
  ``Flavor decomposition of nucleon structure at a neutrino factory,''
  hep-ph/0009223.
  
  \bibitem{Hughes:1983kf}
  For a very early review, see V.~W.~Hughes and J.~Kuti,
  Ann.\ Rev.\ Nucl.\ Part.\ Sci.\  {\bf 33} (1983) 611.
   
\bibitem{Anselmino:1992vg}
  M.~Anselmino, E.~Predazzi, S.~Ekelin, S.~Fredriksson and D.~B.~Lichtenberg,
  Rev.\ Mod.\ Phys.\  {\bf 65} (1993) 1199.


 \bibitem{Bj2010}
J. Bjorken, ``The Parton Model: 2010,'' talk at the Rockefeller Spring Workshop on Electron-Nucleus Collider Physics, http://j.mp/q1fkVA. 

  \bibitem{sCPp} For accessible reviews, see H.~R.~Quinn, ``CP Symmetry Breaking, or the Lack of It, in the Strong Interactions,''
\href{http://j.mp/hKIVLe}{SLAC-PUB-10698 (2004)}; 
M.~Dine, ``The Strong CP Problem,'' in \textit{Flavor Physics for the Millennium,} ed.
J.~L.~Rosner, World Scientific, Singapore, 2000, hep-ph/0011376.
  
  \bibitem{Khachatryan:2010gv}
  V.~Khachatryan {\it et al.}  [CMS Collaboration],
  JHEP {\bf 1009} (2010) 091.


\bibitem{Bjorken:2013boa}
  J.~D.~Bjorken, S.~J.~Brodsky and A.~S.~Goldhaber,
  arXiv:1308.1435.
 
 \bibitem{Quigg:2010nn}
  C.~Quigg, {Nuovo Cim.} \textbf{33C,} 327 (2011);
  ``Learning to See at the Large Hadron Collider,''
  arXiv:1001.2025.


\bibitem{Gross:1973id}
  D.~J.~Gross and F.~Wilczek,
  Phys.\ Rev.\ Lett.\  {\bf 30} (1973) 1343;
    H.~D.~Politzer,
  Phys.\ Rev.\ Lett.\  {\bf 30} (1973) 1346.

 
\bibitem{Kronfeld:2010bx}
 A recent example appears in Figure~3 of  A.~S.~Kronfeld and C.~Quigg,
  Am.\ J.\ Phys.\  {\bf 78} (2010) 1081.
  
  \bibitem{vanRitbergen:1997va}
  T.~van Ritbergen, J.~A.~M.~Vermaseren and S.~A.~Larin,
  Phys.\ Lett.\ B {\bf 400} (1997) 379.


\bibitem{Kronfeld:2012ym}
Hadron mass spectra computed with $2+1$ flavors of sea quarks are displayed in Figure~3 of
  A.~S.~Kronfeld,
  ``Lattice Gauge Theory and the Origin of Mass,'' in \textit{100 Years of Subatomic Physics,} ed. E.~Henley and S.~D.~Ellis (World Scientific, Singapore, 2013) p.~493
  [arXiv:1209.3468].
  
  \bibitem{GaLeu} The line of development that leads to this conclusion begins with J.~Gasser and H.~Leutwyler, Nucl. Phys. B250 (1985) 465. See also N.~Fettes and U.~G.~Meissner, Nucl. Phys. A676  (2000) 311; V.~Bernard and U.~G.~Meissner, Annu. Rev. Nucl. Part. Sci. 57 (2007) 33; S.~R.~Beane and M.~J.~Savage, Nucl. Phys. A717  (2003) 91.
  
  \bibitem{CQPT}
  C.~Quigg, ``Top--ology, Phys. Today 50 (May, 1997) 20 ; extended version circulated as hep-ph/9704332.
  
  \bibitem{Quigg:2012zm}
  C.~Quigg,
  EPJ Web Conf.\  {\bf 28} (2012) 01001.
  
\bibitem{ATLAS:2013lla}
The  ATLAS Collaboration,
  ``Measurement of multi-jet cross-section ratios and determination of the strong coupling constant in proton-proton collisions at $\sqrt{s}=7\tev$ with the ATLAS detector.,''
  \href{http://cds.cern.ch/record/1543225}{ATLAS-CONF-2013-041};
  S.~Chatrchyan {\it et al.}  [CMS Collaboration],
  ``Measurement of the ratio of the inclusive 3-jet cross section to the inclusive 2-jet cross section in $pp$ collisions at $\sqrt{s}$ = 7 TeV and first determination of the strong coupling constant in the TeV range,''   arXiv:1304.7498.
  
\bibitem{gaugecan} See the compilation made by the LEP Electroweak Working Group, \href{http://lepewwg.web.cern.ch/LEPEWWG/}{http://lepewwg.web.cern.ch/LEPEWWG/}.  
  
\bibitem{Hdisc} 
  G.~Aad {\it et al.}  [ATLAS Collaboration],
  Phys.\ Lett.\ B {\bf 716} (2012) 1;
    S.~Chatrchyan {\it et al.}  [CMS Collaboration],
  Phys.\ Lett.\ B {\bf 716} (2012) 30;
  M.~Della~Negra, P.~Jenni, and T.~S.~Virdee, Science \textbf{338} (2102) 1560;
  The CMS Collaboration, Science \textbf{338} (2012) 1569;
  The ATLAS Collaboration, Science \textbf{338} (2012) 1576.  
  
\bibitem{Hmech} F.~Englert and R.~Brout, Phys. Rev. Lett. \textbf{13} (1964) 321;  
P.~W.~Higgs, Phys. Rev. Lett. \textbf{13} (1964) 508; Phys. Lett. \textbf{12} (1964) 132;
G.~S.~Guralnik, C.~R.~Hagen, and T.~W.~B.~Kibble, Phys. Rev. Lett. \textbf{13} (1964) 585.

  
\bibitem{ATLASHiggs} \href{https://twiki.cern.ch/twiki/bin/view/AtlasPublic/HiggsPublicResults}{ATLAS Public Higgs Results;} K.~Jakobs, ``Higgs Boson Physics at ATLAS,'' \href{https://indico.cern.ch/getFile.py/access?contribId=4&sessionId=2&resId=0&materialId=slides&confId=242095}{Report to the XXVI International Symposium on Lepton Photon Interactions at High Energies} (June 2013).

\bibitem{CMSHiggs} \href{https://twiki.cern.ch/twiki/bin/view/CMSPublic/PhysicsResultsHIG}{CMS Public Higgs Results;} A.~DeRoeck, ``Higgs Physics at CMS,'' \href{https://indico.cern.ch/getFile.py/access?contribId=5&sessionId=2&resId=1&materialId=slides&confId=242095}{Report to the XXVI International Symposium on Lepton Photon Interactions at High Energies} (June 2013).

\bibitem{HWG} \href{https://twiki.cern.ch/twiki/bin/view/LHCPhysics/CrossSections}{LHC Higgs Cross Section Working Group.}

\bibitem{Quigg:2009xr}
  C.~Quigg and R.~Shrock,
  Phys.\ Rev.\ D {\bf 79} (2009) 096002.
  
\bibitem{BruceM} See the talk at DIS2013 by B.~Mellado, ``Higgs in $ep$ at the LH$e$C,'' \href{http://j.mp/1dRJg52}{http://j.mp/1dRJg52}.  

\bibitem{Asner:2013hla}
  For an informal review, see M.~Schulze, \href{https://indico.fnal.gov/getFile.py/access?contribId=3&resId=0&materialId=slides&confId=6983}{Top Precision Measurements,} 5th TLEP Workshop, Fermilab, July 2013. For recent work, see
  D.~Asner \textit{et al.}, 
  ``Top quark precision physics at the International Linear Collider,''
  arXiv:1307.8265.
  
\bibitem{ErlLang} Among many reviews, see J.~Erler and P.~Langacker, ``Electroweak model and constraints on new physics'' in J.~Beringer {\it et al.} [Particle Data Group], Phys. Rev. D\textbf{86} (2012) 010001, and 2013 partial update for the 2014 edition.  

\bibitem{Quigg:2009vq}
  C.~Quigg,
  Ann.\ Rev.\ Nucl.\ Part.\ Sci.\  {\bf 59} (2009) 505.
  
\bibitem{Branco} G.~C.~Branco and M.~N.~Rebelo, 
  PoS Corfu {\bf 2012} (2013) 024.

\bibitem{Glashow:1970gm}
  S.~L.~Glashow, J.~Iliopoulos and L.~Maiani,
  Phys.\ Rev.\ D {\bf 2} (1970) 1285.

%
\bibitem{Aaij:2013aka}
  R.~Aaij {\it et al.}  [LHC$b$ Collaboration],
  ``Measurement of the $B^0_s \to \mu^+ \mu^-$ branching fraction and search for $B^0 \to \mu^+ \mu^-$ decays at the LHCb experiment,''
  arXiv:1307.5024.
  S.~Chatrchyan {\it et al.}  [CMS Collaboration],
  ``Measurement of the $B_s \to \mu \mu$ branching fraction and search for $B^0 \to \mu \mu$ with the CMS Experiment,''
  arXiv:1307.5025.
  The CMS and LHC$b$ Collaborations, ``Combination of results on the rare decays $B \to \mu\mu$ from the CMS and LHC$b$ experiments,'' \href{http://cds.cern.ch/record/1564324}{CMS-PAS-BPH-13-007}.
  
\bibitem{Buras:2012ru}
  A.~J.~Buras, J.~Girrbach, D.~Guadagnoli and G.~Isidori,
  Eur.\ Phys.\ J.\ C {\bf 72} (2012) 2172.
  
%
\bibitem{Buras:2013uqa}
  A.~J.~Buras, R.~Fleischer, J.~Girrbach and R.~Knegjens,
  JHEP {\bf 1307} (2013) 77.
  
  %

\end{thebibliography}
\end{document}